\documentclass{aip-book}

\usepackage{graphicx,hologo,mwe,url} 
\usepackage{array,booktabs,threeparttable} 
\usepackage{harvard} 

\usepackage{amsmath}

\usepackage{amsthm}

\theoremstyle{definition}

\usepackage{booktabs}
\usepackage[table,xcdraw]{xcolor}

\usepackage{upgreek}

\begin{document}
\setchapter{1} 
\chapter{Triton and Pluto: same origin but separated at birth} 

\noindent\textbf{Olivier Mousis}\\
Aix-Marseille Université, CNRS, CNES, Institut Origines, LAM, Marseille, France\\
Institut Universitaire de France (IUF)\\
olivier.mousis@lam.fr\\

\noindent\textbf{Sarah E. Anderson}\\
Aix-Marseille Université, CNRS, CNES, Institut Origines, LAM, Marseille, France\\

\noindent\textbf{Adrienn Luspay-Kuti}\\
Johns Hopkins University Applied Physics Laboratory, 11100 Johns Hopkins Road, Laurel, MD 20723, USA\\

\noindent\textbf{Kathleen E. Mandt}\\
NASA Goddard Space Flight Center, Greenbelt, MD, USA\\


\noindent\textbf{Pierre Vernazza}\\
Aix-Marseille Université, CNRS, CNES, Institut Origines, LAM, Marseille, France\\

\begin{abstract}
Assessing the origin of Pluto and Triton has profound implications for the bigger picture of Solar System formation and evolution. In such a context, this chapter reviews our current knowledge of the formation conditions of Pluto and Triton's constitutive building blocks in the protosolar nebula, which can be derived from their known or estimated volatile contents. Assuming that the ultravolatiles carbon monoxide and dinitrogen detected in Pluto and Triton are primordial, the presence of these molecules suggest that the two bodies accreted material originating from the vicinity of the carbon monoxide and dinitrogen icelines. Dinitrogen--rich and water--poor comets such as comet C/2016 R2 (PanSTARRS) obviously present a compositional link with Pluto and Triton, indicating that their building blocks formed in nearby regions of the protosolar nebula, despite of the variation of the water abundance among those bodies. Also, the assumption of Triton's growth in Neptune's circumplanetary disk requires that its building blocks formed at earlier epochs in the protosolar nebula, to remain consistent with its estimated composition.
\end{abstract}


\section{Introduction}

Our knowledge of the current composition of Pluto and Triton provides a starting point for determining where and when each of these bodies formed and how they ended up at their present locations. Assessing their origin has profound implications for the bigger picture of Solar System formation and evolution, particularly the formation locations and the growth timescales of the four giant planets, the extent of their migration after formation, and how this latter shaped the present-day physical and orbital properties of smaller bodies.

Triton and Pluto have long been recognized as twin worlds with similar sizes, densities, and even compositions. While these uncanny similarities provide key constraints on the formation conditions of their building blocks in the protosolar nebula (PSN), the nuances in their compositional differences carry just as much weight. Both bodies have predominantly dinitrogen (N$_2$)--dominated atmospheres with minor amounts of methane (CH$_4$) and carbon monoxide (CO), implying that their building blocks must have formed in nitrogen-rich regions of the PSN \cite**{Br89,Cr93,Le17,Yo18}. The atmospheres are buffered by sublimation equilibrium with some mixture of the same ices (predominantly N$_2$) on the surface \cite**{Ye95,Be16}. Solar ultraviolet flux initiates N$_2$-CH$_4$ photochemistry in both atmospheres, which leads to the production of more complex hydrocarbon species and haze \cite**{Wo17,Be22}. However, there are notable differences between Triton and Pluto when it comes to both the atmospheric abundances of CH$_4$ and CO, and their photochemistry (see Nordheim et al., this book). 

Our knowledge about Triton's and Pluto's surface and atmospheric compositions is skewed toward Pluto thanks to the close flyby by the \textit{New Horizons} spacecraft of the Pluto system in 2015. In contrast, the only spacecraft to ever visit Triton so far was \textit{Voyager 2} back in 1989. While a mission to Triton in the near future is absolutely necessary to further our understanding of this moon, the Neptune system, and potential ocean worlds in general, we are currently limited to the \textit{Voyager 2} observations, and applying lessons learned at Pluto to Triton. 

This chapter reviews our current knowledge of the formation conditions of Pluto and Triton's constitutive building blocks, which can be derived from their known or estimated volatile contents. The known compositions of Pluto and Triton's atmospheric, surface and bulk compositions are presented in Section 2. This section also describes the different mechanisms that may have been at play to shape Pluto and Triton's volatile budget over time. Section 3 establishes connections between the compositions of comets, in particular those of N$_2$--rich bodies such as comet C/2016 R2 (PanSTARRS) (R2), with those of Pluto and Triton. The formation conditions of Pluto and Triton's building blocks in the PSN are also discussed in light of recent works depicting the evolution of the radial abundance profiles of volatiles in the outer regions of the disk. The different scenarios depicting the dynamical origin of Triton are presented in Section 4. Section 5 is dedicated to discussion and conclusion.

\section{Composition of Pluto and Triton}
\label{sec:sec2}

This section reviews the known or estimated abundances of the different volatile reservoirs in both Pluton and Triton. The processes that might have affected the volatile content of the two bodies over time are also discussed.

\subsection{Atmospheric and Surface Composition}

During its flyby, the \textit{New Horizons} spacecraft found that Pluto’s neutral atmosphere consists primarily of more than 99\% N$_2$, $\sim$0.30\% CH$_4$ \cite**{Yo18}, and $\sim$0.05\% CO \cite**{Le17}. The hydrocarbons ethane (C$_2$H$_6$), acetylene (C$_2$H$_2$) and ethylene (C$_2$H$_4$) were also detected below $\sim$500 km, with middle atmospheric mixing ratios of $\sim$0.001 for all three species, dropping to $2\times10^{-5}$, $5\times10^{-6}$ and $6\times10^{-7}$ in the middle atmosphere at $\sim$100 km, respectively \cite**{Yo18}. The near-surface CO/N$_2$ and the CO/CH$_4$ mixing ratios were found to be $\sim$4$\times10^{-3}$ and $\sim$1.7$\times10^{-3}$, respectively \cite**{Le17}. 

On the other hand, the \textit{Voyager 2} flyby showed that Triton's atmospheric CH$_4$ is about one order of magnitude less abundant, compared with the \textit{New Horizons} measurements made at Pluto. \textit{Voyager} observations only provided an upper limit for CO in Triton's atmosphere \cite**{Br89}, but ground-based observations of its surface ice gave a CO/N$_2$ ratio of $\sim$0.1\%, allowing to predict that the atmospheric mixing ratio would be 1.5 x 10$^{-4}$ \cite**{Cr93}. Later ground-based observations detected CO at abundances similar to the surface ice observations \cite**{Le10}. The resulting CO/CH$_4$ ratio in Triton's atmosphere is then $\sim$3.5, roughly three orders of magnitude higher than in Pluto's atmosphere. The resulting CO/N$_2$ ratio in Triton's atmosphere is found to be $\sim$0.015\%, namely a factor of $\sim$6 larger than in Pluto's atmosphere. 

One of the biggest holes in our knowledge about Triton comes from the fact that no instrument on \textit{Voyager 2} measured the composition of its surface. Hence, our information about Triton's surface composition is limited, and solely comes from ground-based spectroscopic observations and modeling efforts. What we do know is that, similarly to Pluto, Triton's surface is dominated by N$_2$ ice, and to a lesser extent CH$_4$ and CO ices \cite**{Ho16}. These ices likely form solid solutions on the surface of Triton, but pure CH$_4$ ice may also form discreet patches. Considering the lack of spacecraft measurements, we have poor knowledge on the distribution of the ices on Triton's surface. Nevertheless, these ices are expected to migrate across the surface in response to the seasonally changing solar insolation \cite**{Cr84,Cr93,Ba10,Bu11}. 

On Pluto, most of the N$_2$ ice seems to be concentrated in the basin called Sputnik Planitia, whereas the CH$_4$ ice is more widely distributed across the surface \cite**{Sc21}. The ice component of the bedrock on both Triton and Pluto includes H$_2$O ice. However, Triton's surface also has detectable amounts of CO$_2$ \cite**{Cr93}, most likely in the form of exposed deposits, whereas this condensate appears to be absent from Pluto's surface. Small amounts of CO ice, H$_2$O ice and NH$_3$ hydrates were also detected on Pluto's surface \cite**{Gr16,Da18,Co19}. Heavier photochemistry products, such as methanol (CH$_3$OH) and hydrocarbon ices are also present on the surface in trace amounts \cite**{Co19}. A layer of heavier hydrocarbons and nitriles on the surface \cite**{Br89} is also possible due to photochemistry, and hazes are observed at low altitudes as well (see Gao et al., this book). The noble gas argon (Ar) was not directly detected by \textit{New Horizons} in Pluto's atmosphere, but an upper limit of 6\% of the column density of CH$_4$ has been estimated \cite**{St20}. Similarly, Ar has also been proposed to be present on Triton, with an upper limit of 10\% of N$_2$'s column density \cite**{McKinnon_NT}.   

\subsection{Bulk Composition}
\label{bulk}

While surface ices and atmospheric compositions are two important reservoirs of volatiles, it is the bulk composition of a planetary body that needs to be assessed for understanding its origin and evolution. The estimates for the current bulk abundances should include volatiles and non-volatiles found on the surface and in the atmosphere, as well as what is predicted to be in the interior. Uncertainties on interior compositions can be large \cite**{Mc19} and the full range of possibilities needs to be considered. 

The bulk amount of H$_2$O can be estimated using the bulk density, which provides some potential information about the fractions of rock and water (ice and potential subsurface ocean) present. The average bulk density of Pluto is $\sim$1854 kg m$^{-3}$ \cite**{Ni17} and the bulk density of Triton is $\sim$2061 kg m$^{-3}$ \cite**{McKinnon_NT}. In the case of Pluto, the water mass fraction in the bulk is between 0.28 and 0.36 \cite{Gl18}. A similar calculation for Triton results in a 0.21--0.28 water mass range in the bulk. Using the masses of Pluto and Triton and converting to moles, the abundance of H$_2$O in their bulks is estimated to be (2--2.58) $\times$ $10^{23}$ moles and (1.5--2.03) $\times$ $10^{23}$ moles, respectively. 

The total amount of N$_2$ in Pluto and Triton is of special importance for constraining their region of formation within the early Solar System, and understanding how closely related their formation was to each other. Whether present-day N$_2$ was originally accreted as N$_2$, NH$_3$, or N-bearing organic material has important implications on the conditions under which these twin-worlds formed (see Sec. \ref{form}). Based on infrared absorption measurements, Pluto's surface N$_2$ reservoir is mainly concentrated to Sputnik Planitia. This reservoir is estimated to outweigh the atmospheric N$_2$ reservoir by orders of magnitude. Thus, taking the surface reservoir to represent the current amount of N$_2$ in the bulk is a reasonable first-order assumption \cite**{McKinnon_2021,Gl18}. The apparent amount of N$_2$ in Sputnik Planitia is $\sim$ (0.4--3) $\times$ $10^{20}$ moles.  Table \ref{tab:N2inventory} summarizes the estimated moles of N$_2$ and H$_2$O in the bulk composition of Pluto and Triton.

\begin{table}[]
\caption{Estimated moles of N$_2$ and H$_2$O in the bulk composition of Pluto and Triton. The surface inventory is assumed to be representative of the inventory in the bulk.}
\label{tab:N2inventory}
\begin{tabular}{@{}lll@{}}
\toprule
       & Moles of N$_2$            & Moles of H$_2$O            \\ \midrule
Pluto   & $(0.4-3)\times 10^{20}$   & $(2-2.58)\times 10^{23}$   \\
Triton & $(0.7-1.4)\times 10^{21}$ & $(1.5-2.03)\times 10^{23}$ \\ \bottomrule
\end{tabular}
\end{table}

Unfortunately, no absorption measurement is available for Triton's surface, but the volatile abundances can be estimated from existing observations. Considering the \textit{Voyager 2} atmospheric measurements, the energy-limited mass flux of N$_2$, and some limits to the thickness of N$_2$ polar frost deposits, an upper limit of 0.5 to 1 km thickness of N$_2$ averaged over the surface of Triton is obtained \cite{McKinnon_NT}. Using these values, the amount of N$_2$ on Triton is estimated to be $(0.7-1.4)\times 10^{21}$ moles. 

The second most abundant component, CH$_4$, has a mole fraction of (3--3.6) $\times$ $10^{-3}$ \cite**{Protopapa_2017} on Pluto's surface diluted in N$_2$. CH$_4$ is predicted to be incorporated in the N$_2$ ice of the surface at a mole fraction of $\sim$ (1--5 ) $\times$  $10^{-4}$ on Triton's surface, yielding a global equivalent layer of $\sim$1 meter \cite**{Mandt23}. The CO mole fraction relative to N$_2$ is  $\sim$(2.5--5) $\times$ $10^{-3}$ on Pluto's surface \cite**{Ow93,Me15}. The Pluto CO/N$_2$ ratio is about six orders of magnitude lower than the ratio measured in comet 67P/Churyumov-Gerasimenko ($\sim$35; \citeasnoun**{McKinnon_2021}). Estimates for Triton's CO/N$_2$ surface mole fraction may vary between $<$0.01 to 0.15, depending on the atmospheric CO/N$_2$ estimate considered, and whether one assumes CO is mixed in the ice, or if it is physically separate from the N$_2$ ice. Table \ref{inventory} summarizes the mole fractions of the most abundant volatile species, relative to N$_2$, in the total inventory of Pluto and Triton.

\begin{table}[]
\caption{Mole fractions of the most abundant volatile species, relative to N$_2$, in the total inventory of Pluto and Triton. }\label{tab:ratios}
\begin{tabular}{@{}lcl@{}}
\toprule
        & \multicolumn{2}{c}{N$_2$}                                                                      \\ \midrule
        & Pluto                                          & \multicolumn{1}{c}{Triton}                    \\ \midrule
H$_2$O  & \multicolumn{1}{l}{(0.7--6.5) $\times$ $10^{3}$}      & $<$107.2--291.5 \\
CH$_4$  & \multicolumn{1}{l}{(3--3.6) $\times$ $10^{-3}$}    & $\sim$ (1--5)  $\times$ $10^{-4}$                   \\
CO      & \multicolumn{1}{l}{ (2.5--5) $\times$ $10^{-3}$}   &   $<$ 0.01--0.15                            \\
CO$_2$  & not detected                                              & (1.5--100) $\times$ $10^{-3}$                     \\ 
\bottomrule
\end{tabular}
\label{inventory}
\end{table}

\subsection{Time evolution of the volatile budget in Pluto and Triton} 
\label{budget}
The process for connecting the current volatiles of Pluto and Triton to their original building blocks begins with assessing the current bulk volatile composition of each body as described in Sec. \ref{bulk}. The measurements that are most useful for studying formation and evolution are noble gas abundances, noble gas isotopes, abundances of carbon, nitrogen, and oxygen, and their isotopes. Unfortunately, we currently only have limited information about carbon, nitrogen, and oxygen for Pluto and Triton, and only an upper limit on argon and on nitrogen isotopes at Pluto.  

After determining the current composition, the next step is to evaluate what processes will influence volatiles in a way that changes the bulk composition of each body. Pluto and Triton have both experienced many processes that can change the composition of their volatiles, both during formation and following it. These processes will convert molecules from one form of volatile to another, like NH$_3$ to N$_2$, but will have no impact on the bulk elemental composition of volatiles in either body. The simplest approach, given such limited information, is to evaluate volatile abundances in terms of relative elemental abundances, such as N/C and O/C. This would enable a comparison with measurements of building block analogs like comets and chondrites, and models for icelines that take into consideration various forms of ices (see approach from \citeasnoun**{Ma22} for Lunar volatiles). We outline below the processes thought to have affected the volatile composition at Pluto and Triton and discuss whether they will have an impact on bulk elemental composition over time. 

The earliest process would be accretional heating from impacts that take place during formation. Accretional heating could cause vaporization of hypervolatiles like N$_2$, CO, and CH$_4$ \cite**{Mc19} leading to the formation of an early atmosphere. Any potential loss would be through atmospheric processes that are discussed later. The possible giant impact that formed the Pluto-Charon system \cite**{Ca21}, and any impact that took place as part of Triton's capture by Neptune \cite**{Ru17} could also have caused heating \cite**{Ca21} that could contribute to the formation of an early atmosphere and atmospheric loss processes. Once formation is complete, impacts would then add volatiles, including hypervolatiles, to the atmosphere \cite**{Si15}. These processes would not only affect the bulk composition of the molecules present, but could also change the relative elemental composition by adding or removing nitrogen, carbon, oxygen, and hydrogen. 

After formation, differentiation of the interior is thought to have led to the formation of a subsurface ocean where hydrothermal processes could have changed the composition of the molecular species that were originally accreted \cite**{Sh93,Gl18}. These processes depend on the internal temperature and pressure, the pH in the water, the oxidation state of the system, and the bulk abundance of nitrogen and carbon. Of particular interest for Pluto and Triton, because of the presence of N$_2$, CH$_4$, and CO on the surface and in the atmosphere, are the conversions between N$_2$ and NH$_3$, and between CO$_2$ and CH$_4$ \cite**{Gl08}. Formation of N$_2$ and CO$_2$ tend to go together and are preferred at higher temperatures, lower pressures, and with systems that are more oxidized and have higher bulk abundances of nitrogen \cite**{Gl08}. Additionally, CO can also be converted to CO$_2$ or CH$_4$ through aqueous chemistry in the interior \cite**{Gl17}. These processes change the molecular composition of the volatiles that are present, but will not affect the bulk elemental composition, such as N/C. In addition to the  conversion reactions, hydrothermal reactions on early Pluto/Triton could have led to a loss of reactive volatiles (e.g., NH$_3$) at 100--300 $^{\circ}$C, forming complex organic molecules \cite**{Se17}.
When the rocky core temperature becomes high, e.g., $>$ $\sim$500 $^{\circ}$C, organic matter contained in the rocky core could also have been thermally decomposed via a suite of irreversible reactions, possibly providing large amounts of volatiles, such as CO$_2$, CH$_4$, and N$_2$ \cite**{Ok11,Re23}. There are however aqueous reactions that can lead to permanent loss of CO \cite**{Sh93,Ne15,Gl17}, which would change the bulk elemental ratios from what existed in the building blocks to values with reduced amounts of carbon and oxygen. 

The surface compositions of Pluto and Triton, potentially provide some constraints on the effectiveness of these hydrothermal processes \cite**{Mandt23}. Both Pluto and Triton have N$_2$, CH$_4$, and CO molecules on the surface and in the atmosphere. The presence of CH$_4$ suggests that production of N$_2$ and CO$_2$ was not complete in the interior unless all of the CH$_4$ currently on the surface was delivered through later impacts \cite**{Si15}. Similarly, the presence of CO suggests either that any conversion of CO to CO$_2$ or CH$_4$ was also not complete unless all of the CO currently observed was also delivered through later impacts \cite**{Gl17}. Triton has CO$_2$ on the surface, but none has been detected on Pluto \cite**{Ah22}, which might suggest that internal chemistry on Triton was more effective than at Pluto \cite**{Mandt23}. Additionally, NH$_3$ has been detected on the surface of Pluto further supporting the idea that any internal conversion to N$_2$ was not complete \cite**{Gl17,Mc19,Mandt23}. Although NH$_3$ has not been detected on the surface of Triton, this may be due to the lack of high spatial and spectral resolution observations like the ones made by New Horizons at Pluto. 

Two additional processes can remove molecules from the observed inventory without removing them from the bulk composition. The first is sequestration of molecules in clathrates in the subsurface ocean. Clathrates are water-ice cages that trap other molecules and atoms. They are proposed to have formed at the base of Pluto’s ice shell allowing the ice shell to harden while an ocean continues to exist thanks to the heat retained by the insulating layer of clathrates \cite**{Ka19}. Although N$_2$, CO, CO$_2$, and CH$_4$ can all be trapped in clathrates, CO clathrate is more stable than N$_2$ clathrate. Also, CH$_4$ and CO$_2$ clathrates are more stable than CO clathrate \cite**{Mc19}. This means that any of the carbon-bearing molecules produced in the interior through aqueous chemistry could be trapped in the interior long term in a layer of clathrates and appear to be missing from the bulk carbon inventory. The other process is burial of CO ice under N$_2$ ice in locations like Sputnik Planitia on Pluto \cite**{Gl17}. This would remove CO, and potentially CH$_4$, from the observed inventory reducing the apparent bulk carbon abundance.

Finally, atmospheric processes of photochemistry and escape will change the molecular composition and can lead to permanent loss of some volatiles through haze formation and by loss from the top of the atmosphere \cite**{Ma17}. Photochemistry will remove N$_2$ and CH$_4$ by converting them to larger organic molecules that eventually form haze \cite**{Lu17,Ma17,Wo17}, similar to processes on Titan \cite**{Ma09,Ma14}. This would reduce the observed N and C abundances over time, but is limited by the number of photons able to reach Pluto and Triton. Escape can occur through thermal processes where heating of the upper atmosphere gives molecules and atoms enough energy to escape the gravity of a planet. It can also occur through ionization and removal of ions by pickup processes in the solar wind or Triton’s magnetosphere. In the case of the atmospheres of Pluto and Triton, both photochemistry and escape would preferentially remove CH$_4$ \cite**{Mandt23}. 

\begin{figure*}
\includegraphics[angle=0,width=14cm]{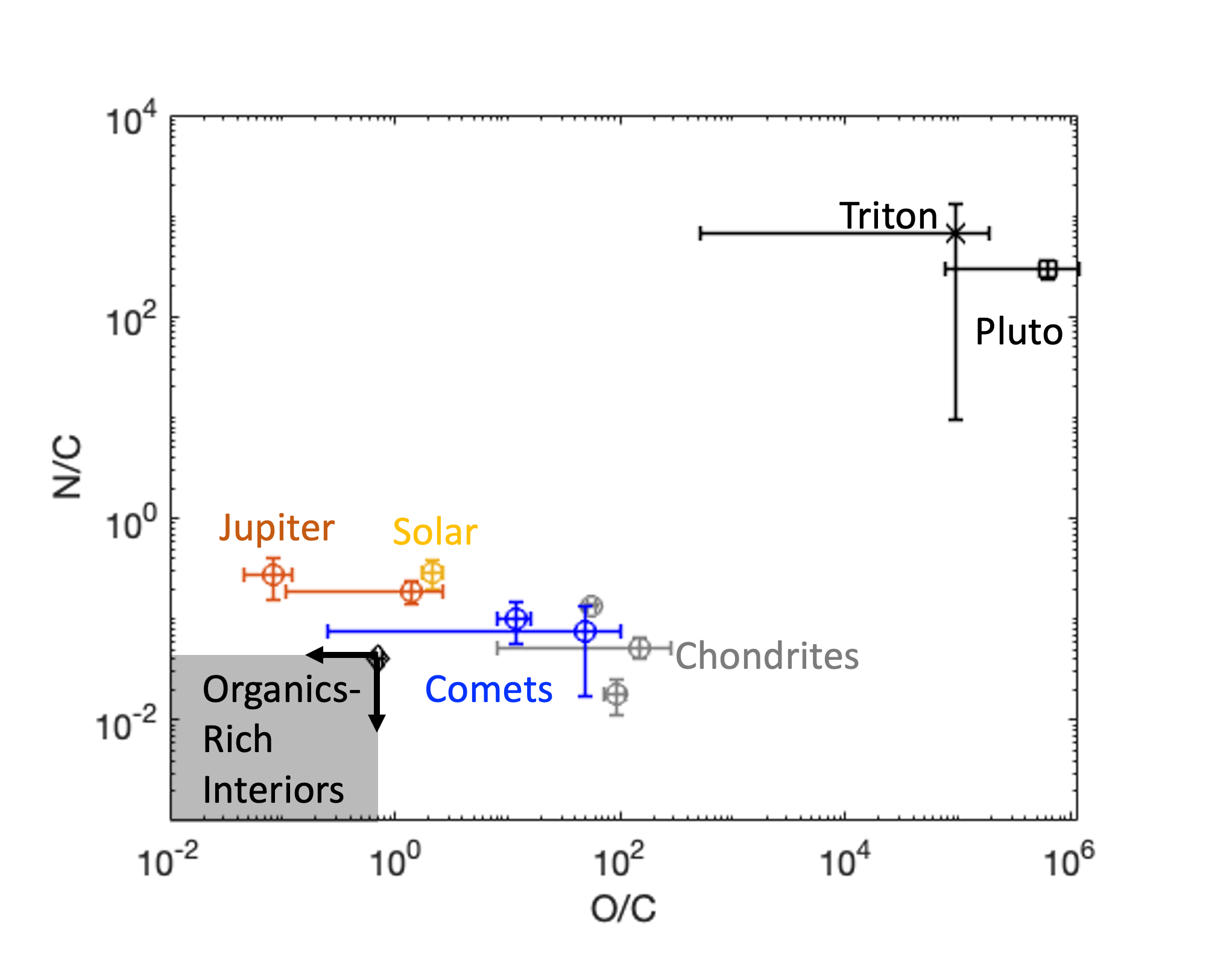}
\caption{Elemental ratios providing information on the relative carbon abundance observed for Pluto and Triton. These ratios are compared to the solar abundances, abundances observed in Jupiter’s atmosphere, and analogs for solid materials in the PSN. The two measurements for Jupiter were made by the Galileo Probe Mass Spectrometer (GPMS) and the Juno Microwave Radiometer (MWR).}
\label{fig_CD}
\end{figure*}

A recent review of the elemental ratios found that the observed volatiles of Pluto and Triton are carbon-poor \cite**{Mandt23} as shown in Fig. \ref{fig_CD}. The only way to produce these observations is through the removal of carbon by preferentially removing CH$_4$ through atmospheric processes, a very reasonable possiblity given that CH$_4$ would be the lightest main species in the atmosphere and the most easily removed \cite**{Mandt23}. This study also noted that the upper limit for Ar in Pluto's atmosphere based on New Horizons observations provides a lower limit for N/Ar that is much larger than the solar value. Because atmospheric loss processes would favor loss of nitrogen over argon, any primordial ratio is likely to be even larger than the lower limit suggesting that the nitrogen at Pluto originated as NH$_3$ or organics \cite**{Mandt23}.  

\section{Formation of Pluto and Triton's building blocks in the PSN}
\label{form}

In this section, we draw connections between the compositions of comets, in particular those of N$_2$--rich bodies such as comet C/2016 R2 (PanSTARRS) (R2), with those of Pluto and Triton. The PSN conditions at which the building blocks of Pluto and Triton could have formed are also discussed.

\subsection{Link with atypical comets}

While Pluto and Triton exhibit N$_2$-rich surfaces \cite**{Cr93,Ow93,Qu99,Me18}, bizarrely, comets are usually depleted in this molecule \cite**{Co00}. As cometesimals are thought to have formed from the materials available to them in the PSN at the location of their birth and are relegated to the outer Solar System for the majority of their lifetimes, they preserve the composition of the disk at that precise area and moment in time: a core sample of the PSN, making them among the most pristine bodies currently populating our Solar System. However, due to their inherent dynamical instability, it is impossible to trace a comet's dynamical history back to its formation location. Their ice-rich composition only shows that they formed in the outer parts of the Solar System, beyond Jupiter.

For the most part, they have compositions similar to predicted protosolar abundances, with no depletion of carbon or oxygen, but a clear deficiency of nitrogen \cite**{Ge87}. For example, Comet 1P/Halley has a N/O elemental abundance depleted by a factor of 3 with respect to the solar abundance \cite**{Je91}, and Comet C/1995 O1 (Hale-Bopp) has an inferred N/O elemental depletion of 15 in the gas phase \cite**{Bo00}. This low abundance ratio is attributed to a depletion of N$_2$, the least reactive of all N-bearing species and believed to be the main carrier of nitrogen in the PSN \cite**{Fe04}. Laboratory studies show that ices incorporated into comets at around 50K would have N$_2$/CO $\approx$ 0.06 if N$_2$/CO is $\approx$ 1 in the solar nebula \cite**{Ow95}. However, the typical N$_2$/CO ratio for most comets is $< 10^{-3}$, which is much lower than expected \cite**{Co00}. \citeasnoun**{Mo12} propose that the nitrogen deficiency in comets may be explained through two possible formation conditions: First, comets could have formed under colder conditions ($\leq20$ K), thus incorporating N$_2$; the nitrogen would then disappear due to subsequent internal heating from the decay of radiogenic nuclides. Alternatively, comets could have formed under conditions warmer than 20 K, thereby circumventing the trapping of N$_2$. This dichotomy points towards the necessity of low temperatures ($\sim$20K) for the formation of pure N$_2$ condensate and for the inclusion of N$_2$ in comets.

Recently, long-period comet R2 was revealed to be a CO-rich comet \cite**{Wi18} and strongly depleted in water, with a H$_2$O/CO ratio of 0.0032 \cite**{McK19} with an upper limit of $<$~0.1 \cite**{Bi18}. The spectrum was dominated by bands of CO$^+$ as well as N$_2^+$, the latter of which was rarely seen in such abundance in comets \cite**{Co18,Op19}. It was also found to be both CN-weak and dust-poor \cite**{Op19}. This CO-rich and water-poor composition, along with none of the usual neutrals seen in most cometary spectra, makes R2 a unique and intriguing specimen. The observed emission fluxes have been used to calculate ionic ratios of N$_2^+$/CO$^+$ in the coma of 0.09 \cite**{An22}, the highest of such ratios observed for any comet so far with high-resolution spectroscopy and in line with the predictions of \citeasnoun**{Ow95}. This is larger than the best measurement in comet 67P/Churyumov-Gerasimenko, with an N$_2$/CO ratio of $\sim$2.87 $\times$ 10$^{-2}$ \cite**{Ru20}, though these measurements were obtained much closer to the nucleus using a mass spectrometer. To account for the high N$_2$/CO and CO/H$_2$O ratios measured in R2, it has been proposed that its precursor grains condensed in the vicinity of the CO and N$_2$ icelines \cite**{Mo21}. This would indicate that R2--like comets formed in a colder environment than the other comets sharing more usual compositions. 

Despite its H$_2$O deficiency, the CO- and N$_2$-rich composition of comet R2 bears a closer resemblance to the compositions of Pluto and Triton than to those of any other comet observed thus far, hinting at the possible formation from similar building blocks (see Sec. \ref{role}). Nevertheless, there are discrepancies when comparing R2's relative abundances of CO, CH$_4$, and N$_2$ to the surface spectra of Pluto. These inconsistencies may be attributed to variations in the surface compositions of large KBOs, as their surfaces are likely modified by the processes mentioned in Sec. \ref{budget}. Additionally, \citeasnoun**{Bi18} suggested that R2 could be a fragment of a differentiated KBO object, also supported by its dust-poor composition. \citeasnoun**{De21} proposed that R2 could be a nitrogen iceberg, a fragment of a differentiated KBO surface formed during periods of dynamical instability. While this hypothesis presents an intriguing explanation for R2's composition, it is challenged by the difficulty of preserving N$_2$ content during the energetic collisions necessary to create such fragments \cite**{Le21}. Consequently, although R2's N$_2$-rich composition provides compelling insights, the mechanisms of its N$_2$ retention, as well as its precise formation conditions, warrant further exploration in future studies.

\subsection{Role of CO and N$_2$ icelines}
\label{role}

The Solar System began as a cloud of interstellar gas and micrometric dust particles, which collapsed to form a protoplanetary disk (PPD), commonly named the PSN. Within such a gas-dominated PPD, dust grains stuck together to form pebbles in the mm-cm range. The difference in the motions of the gas and solid particles in the PPD led to streaming instabilities that locally eased the formation of planetesimals \cite**{Yo05,Jo12,Jo15}. These kilometer-sized objects were the building blocks of the Solar System, some growing large enough for their gravity to shape them into spheres, becoming planetary embryos. While gas was still in the disk, these planetary embryos became cores of the giant planets, which were embedded in the disk and could migrate, leading to dynamical instability. Once the gas in the disk was depleted, these went on to become the basis of the formation of the rocky planets, as well as dwarf planets, and large moons.

The composition of available planetesimals was determined by their proximity to specific icelines in the disk. An iceline (also ice line, snowline, frostline, or simply condensation line) is defined as the distance where the surface density of vapor of a given species is equal to that of its solid form in PPDs, the PSN, and circumplanetary disks where satellites form. Inside the snowline, water ice evaporates into water vapor. Outside the snowline, ice is present due to the condensation of vapor, though the motion of particles within the disk allows for solids to exist in front of this line as well as some vapors to exist beyond, due to the kinetics of condensation/sublimation. 

The location of early icelines determined the distribution of volatile species in the disk, which in turn affected the formation of planetesimals and the building blocks of the planets. Factors such as the luminosity of the Sun, the size of dust grains, and the turbulence of the disk also affected the location of icelines in the PSN. The luminosity of the Sun played a critical role because it determined the temperature of the disk at different distances. As the disk evolved over time, so too have the positions of the icelines in our Solar System. This evolution can be caused by various factors such as disk dissipation \cite**{Mo20}, the presence of an inner gap \cite**{Li21}, and interplay between planet formation, pebble accretion and disk evolution \cite**{Bi19}. The size of dust grains also played a role because smaller grains could be more easily lifted by the turbulence of the disk, affecting the transport of volatile species and altering the location of icelines. Studies of the asteroid belt show that the location of the water iceline played a crucial role in determining the distribution of hydrated minerals within the belt \cite**{Ri03}. Similarly, the location of the CH$_4$, CO, and N$_2$ icelines had significant impacts on the formation and composition of the outer planets and their moons \cite**{Ag22,An21}. The interaction of icy pebbles and their vapors around icelines can lead to significant changes in the composition of the PSN, creating a compositional gradient that may be responsible for the volatile enrichment observed in the giant planets of our Solar System \cite**{Mo15,Bo17,De17,Mo19,Ag22}. 

\begin{figure*}
\includegraphics[angle=0,width=8cm]{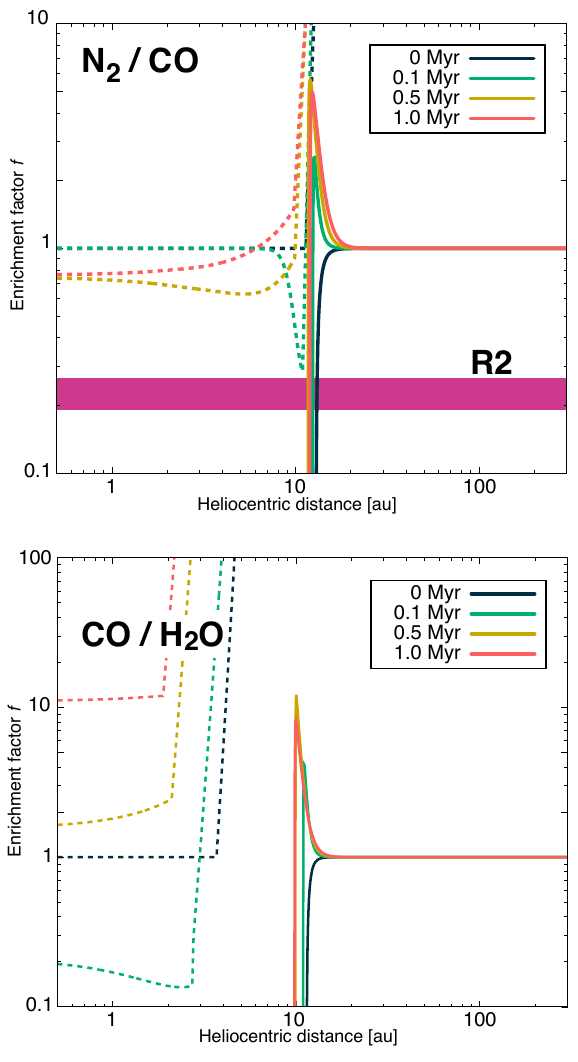}
\caption{Radial profiles of the N$_2$/CO and CO/H$_2$O ratios relative to their initial abundances (defined by the enrichment factor $f$) calculated as a function of time in the PSN for a viscosity parameter $\alpha$~=~$\times$ 10$^{-3}$. Dashed and solid lines correspond to vapor and solid phases, respectively. The purple bar corresponds to the N$_2$/CO ratio measured in comet R2.}
\label{fig_PT}
\end{figure*}

Figure \ref{fig_PT} represents the time evolution of the radial profiles in the PSN of the N$_2$/CO and CO/H$_2$O ratios relatives to their initial abundances, defined by the enrichment factor $f$, in both solid and gas phases, and a value of the viscosity parameter $\alpha$ set to 10$^{-3}$, based on the disk and transport model of \citeasnoun**{Mo21} to which the reader is referred for full details. The adopted $\alpha$ value is well within the range of those typically used in models of PPDs \cite**{He01,Ne13,Si15}. The figure shows that $f$ is flat almost everywhere in the PSN, except in the vicinity of the CO and N$_2$ icelines, which are located at very close distances from each other, i.e. less than a few tenths of AU, in the 10-15 AU region of the PSN, with the CO iceline situated a bit closer in in the disk.

The figure shows that the N$_2$/CO ratio increases to more than $\sim$10 in the gas phase in the area of the two icelines. This peak corresponds to the supply of N$_2$ vapor when N$_2$-rich dust drifts inward the N$_2$ iceline, which is in excess compared to the CO vapor supplied via backward diffusion beyond the CO iceline. When significant, this peak is preceded by a decrease of the N$_2$/CO ratio, which corresponds to the supply of CO vapor when CO-rich dust drifts inward the CO iceline. The N$_2$/CO ratio in solid phase also experiences important depletions that can reach several orders of magnitude, which correspond to the location where N$_2$ essentially forms vapor while CO remains in solid phase. This calculation shows it is possible to form dust in a narrow region of the PSN, i.e. within 10--15 AU, with N$_2$/CO ratios that not only can match the value estimated in comet R2 but also those estimated in the bulk inventory of Pluto and Triton (see Table \ref{inventory}).  On the other hand, when approaching the CO iceline, the CO/H$_2$O ratio varies over several orders of magnitude. It forms a peak at the location of the iceline whose magnitude depends on the adopted $\alpha$--value in the disk \cite**{Mo21}, as a result of the backyard diffusion and condensation of the CO vapor. If the building blocks of R2 were assembled from such grains, they should present a CO--rich and H$_2$O--poor composition, in agreement with the observations. However the CO/H$_2$O ratio is deeply depleted in the solids at distances inward of the CO iceline because this species is in gaseous form. Solids present at very close locations inward the CO iceline would then present CO/H$_2$O ratios compatible with those estimated for Pluto and Triton. 

As mentioned above, the heliocentric distances of the iceline locations are purely indicative as they remain strongly model-dependent. Also, the thermodynamic data for pure condensates should be considered with caution at low pressure conditions, as they have been extrapolated from laboratory data that are several orders of magnitude higher \cite**{Fr09}.

\section{Dynamical origin of Triton}

Triton is unique among the large moons of the Solar System in that it moves in a retrograde and highly inclined - albeit circular - orbit around Neptune. 

The origin of Triton is still an open question, with the two competing hypotheses being 1) the capture of Triton from an heliocentric orbit \cite**{Mc84,Mc95,Ag06,Vo08,No11} and 2) an in-situ formation in the circumplanetary disk (CPD) of Neptune \cite**{Ha79,Sz18,Li20}. In either case, the extraordinary orbital properties of Triton may be a natural consequence of the formation and early dynamical evolution of the outer Solar System. 

Even if the ultimate proof is still missing, there is some level of consensus in the planetary science community regarding the fact that the giant planets may have formed at different heliocentric distances from those where they are presently observed. The idea of a radial migration of the giants planets --and in particular that of Uranus and Neptune-- in the early Solar System \cite**{Fe84,Ma93,Ma95} served as a foundation for many subsequent works including the Nice model \cite**{Ts05}. In this model and its subsequent evolutions \cite**{Ne12,Ne18}, the giant planets are assumed to have formed in a compact configuration (all were located between $\sim$5 and $\sim$15 AU from the Sun), to be possibly more numerous (five instead of today's four; \citeasnoun**{Ne11}, \citeasnoun**{Ne12}) and to be surrounded by a planetesimal disk. Eventually, the orbits of the giant planets became unstable. Uranus and Neptune were gravitationally scattered outwards, thereby penetrating the outer disk and scattering its constituents throughout the Solar System. In this scenario, Triton and Pluto may have a similar origin in the primordial outer disk with Triton being captured by Neptune via gas drag \cite**{Mc84,Mc95} or three-body gravitational interaction \cite**{Ag06,Vo08,No11}. In this latter case, Triton's capture efficiency has been estimated to be between 2\% \cite**{Vo08} and 50\% \cite**{No11}.

The Nice model scenario may also be compatible with an in situ formation of Triton, but requires the existence of five giant planets in the early Solar System as advocated by some versions of the Nice model (see above). The fifth giant planet, before its ejection from the Solar System, probably encountered other planets as well. This might have been the case for Neptune, with the close encounter causing a drastic flip to Triton's initial prograde, circular and equatorial orbit \cite**{Li20}. \citeasnoun**{Li20} found that the encounter with the fifth giant planets injects a large moon onto a retrograde orbit with specific angular momentum similar to Triton’s in 0.3--3\% of their simulations, which is an order of magnitude less efficient than the above capture scenario.

One should note that, if Triton formed in Neptune's CPD instead of the outer PSN, the presence of important amounts of N$_2$ and CO in its interior indicates that accretion from building blocks originating from the PSN instead of having condensed in the CPD itself, assuming those two species are primordial. This would imply that the moons formed in a cool and late CPD around Neptune. In contrast, if CH$_4$ and NH$_3$ were the dominating N-- and C--bearing species in Triton, this would have indicated that their building blocks condensed in a warm and dense CPD \cite**{Pr81}. This implies that, whatever the considered formation scenario is, accretion of Triton in the outer PSN or in Neptune's CPD, its building blocks always originated from the PSN.

\section{Conclusion}

Assuming that CO and N$_2$ have been captured by Pluto and Triton during their formation, the presence of large amounts of these molecules suggest that the two bodies accreted material originating from the outer regions of the PSN. Specifically, volatile transport/condensation models of the PSN suggest that the formation of Pluto and Triton's building blocks in the vicinity of the N$_2$ and CO icelines would be compatible with their present-day compositions. With its CO-- and N$_2$--rich composition, and despite a significant H$_2$O deficiency compared with Pluto and Triton, the unusual comet R2 obviously presents a compositional link with the two bodies, indicating that their building blocks formed in nearby regions of the PSN. Those regions, located around the N$_2$ and CO icelines present substantial variations of the N$_2$/H$_2$O and CO/H$_2$O ratios in the condensed solids and could encompass the peculiar compositions attributed to R2, Pluto, and Triton. Also, the possible formation of Triton in Neptune's CPD remains consistent with its current composition, provided that its building blocks formed before in the PSN.

On the other hand, nothing warrants that Pluto and Triton's current volatile budgets are primordial, as many post-formation processes  could have shaped their current compositions. This implies that any conclusion regarding the origin of the two bodies, and based on their current compositions, must be taken with caution. 

To assess the origin of Pluto and Triton's building blocks, a measurement of the deuterium--to--hydrogen ratio in their H--bearing volatiles would be needed. The comparison of such a value with those measured in comets and the ice giants Uranus and Neptune would put a useful context to the formation of Pluto and Triton's building blocks in the PSN. Measurements of the $^{14}$N/$^{15}$N ratio in the N$_2$ present in Triton and Pluto would be useful as well. Comparing this value with those measured in comets ($\sim$127; \citeasnoun**{Ro14}), solar wind (441; \citeasnoun**{Ma11}) or Titan (143; \citeasnoun**{Ma09}) would provide hints about the primordial form under which nitrogen was accreted by the two bodies, as the $^{14}$N/$^{15}$N ratio in primordial NH$_3$ is expected to strongly depart from the one in primordial N$_2$ \cite**{Ro14}.

\bibliographystyle{dcu}
\bibliography{aipsamp}

\end{document}